\documentclass[iopart,jpb]{iopart}
\usepackage{iopams}
\usepackage[varg]{txfonts}
\usepackage{shortvrb,psfrag,graphicx}
\def\be{\begin{equation}}
\def\ee{\end{equation}}
\def\bea{\begin{eqnarray}}
\def\eea{\end{eqnarray}}

\def\a{\alpha}

\def\o{\omega}

\begin{document}
\title{Exact solutions of the modified Gross-Pitaevskii equation in
`smart' periodic potentials in the presence of external source}
\author{Thokala Soloman Raju$^1$
and Prasanta K Panigrahi$^2$}
\address{
$^1$15647, Avenida, Alcachofa, San Diego, CA 92128, USA \\
$^2$Indian Institute of Science Education and Research (IISER),
Salt Lake, Kolkata 700106, India}
\ead{solomonr\_thokala@yahoo.com,~~prasanta@prl.res.in}

\begin{abstract}
We report wide class of exact solutions of the modified
Gross-Pitaevskii equation (GPE) in `smart' Jacobi elliptic
potentials: $V(\xi)=-V_{0}{\rm sn(\xi,m)}$, $V(\xi)=-V_{0}{\rm
cn(\xi,m)}$, and $V(\xi)=-V_{0}{\rm dn(\xi,m)}$ in the presence of
external source. Solitonlike solutions, singular solutions, and
periodic solutions are found using a recently developed fractional
transform: $\rho(\xi)=\frac{A+Bf^2}{1+Df}$, where $f$ is the
respective Jacobi elliptic function and the amplitude parameters
$A$, $B$, and $D$ {\it nonzero}. These results generalize those
contained in (Paul T, Richter K and Schlagheck~~P 2005 \emph{Phys.
Rev. Lett.} {\bf 94}, 020404) for nonzero trapping potential.
\end{abstract}

\maketitle

\section{Introduction}
In a mean-field approximation, the dynamics of a dilute-gas
Bose-Einstein condensate (BEC) can be captured by the cubic
nonlinear Schr\"odinger equation (NLSE) with a trapping potential
\cite{dalfovo,carr1,bongs}. The various traps which are used to
contain the BEC have spurred the solutions of the NLSE with new
potentials \cite{kunze,carr2}. We consider the mean-field model of
a quasi-one-dimensional BEC trapped in a `smart' potential in the
presence of an external source\cite{paul}
  \be i\frac{\partial\psi}{\partial
t}=\left[-\frac{1}{2}\frac{\partial^{2}}{\partial
x^{2}}+V(\xi)+g\mid\psi\mid^{2}\right]\psi+K{\rm
exp}(i\chi(\xi)-i\o t),\ee where $\psi(x,t)$ represents the
macroscopic wave function of the condensate and $V(\xi)$ is an
experimentally generated macroscopic potential. The parameter $g$
indicates the strength of atom-atom interactions and it alone
decides whether Eq. (1) is attractive ($g=-1$, focussing
nonlinearity) or repulsive ($g=1$, defocussing nonlinearity).
Here, $K$ and $\o$ are real constants related to the source
amplitude and the chemical potential, respectively, and
$\chi(\xi)$ is a real function of $\xi =\a(x-vt)$,  $\a$ and $v$
being two real parameters. In the field of nonlinear optics, Eq.
(1) may describe the evolution of the local amplitude of an
electromagnetic wave in the spatial domain, in a two-dimensional
waveguide where $t$ becomes the propagation distance and $x$
becomes the retarded time, and the system is driven by an external
plane pump wave. The Jacobi elliptic potential may describe a
transverse modulation of the refractive index in the waveguide.

As is well-known, Eq. (1) is not integrable if $KV_{0}\neq 0$, and
only small classes of explicit solutions can most likely exist.
For $V(\xi)=0$, Eq. (1) is a cubic NLSE with a source, and exact
rational solutions using a fractional transform are found in Ref.
\cite{panigrahi}. And in Ref. \cite{bronski3} periodic solutions
without source have been reported. More recently, in
Ref.\cite{kengne}, a class of exact solutions of Eq. (1) for
$V(\xi)=-V_{0}{\rm sn}^{2}(\xi,m)$ have been reported. In
particular, the rational solutions of the fractional transform:
$\rho(\xi)=\frac{A+Bf^{2}}{1+Df^{2}}$, where $f={\rm sn}$ have
been reported for $B=0$. This is due to the form of the trapping
potential. Nonetheless, in the present paper we find rational
solutions of the type $\rho(\xi)=\frac{A+Bf^{2}}{1+Df}$, where $f$
being the respective Jacobi elliptic function with all the
amplitude parameters $A$, $B$, and $D$ {\it nonzero}. These
results generalize those contained in Ref \cite{paul} for nonzero
trapping potential. The choice of a 'smart' potential $V(\xi)$
allows one to construct a large class of exact solutions, as done
in a number of works for the cubic GP equations
\cite{bronski1,bronski2,bronski3}. In the present work, we
consider three different potentials in the GP equation:
$V(\xi)=-V_{0}{\rm sn(\xi,m)}$, $V(\xi)=-V_{0}{\rm cn(\xi,m)}$,
and $V(\xi)=-V_{0}{\rm dn(\xi,m)}$ in the presence of external
source, and find exact travelling wave solutions of Eq. (1) with
$K\neq 0$. The choice of these three different `smart' potentials
is motivated by the following facts. Firstly, the potential
$V(\xi)=-V_{0}{\rm sn(\xi,m)}$ in the limit $m\rightarrow0$ is
$V(\xi)=-V_{0}{\rm sin(\xi)}$ which is similar to the standard
optical lattice potential \cite{Choi,berg}. Secondly, the choice
of the potential $V(\xi)=-V_{0}{\rm dn(\xi,m)}$ in GP equation
mimics\cite{kengne1} the harmonic potential that was used to
achieve BEC experimentally. The third potential, we hope it is
relevant to the available experimental conditions to achieve BEC.
\section{EXACT SOLUTIONS OF THE GPE IN `SMART' PERIODIC POTENTIAL WITH SOURCE}
The travelling wave solutions of Eq. (1) with potential $V(\xi)$
are taken to be of the form
$\psi(x,t)=\rho(\xi)e^{i\chi[\a(x-vt)]-i\o t}$. Inserting this
expression for $\psi(x,t)$ in Eq. (1) and separating the real and
imaginary parts, and integrating the imaginary part, one gets \be
\chi^{\prime}=\frac{v}{\a}+\frac{C}{\a\rho^{2}}, \ee
 where $C$ is a constant of integration. In order that the
external phase be independent of $\psi$, we consider only
solutions with $C=0$, to obtain \be
 \rho^{\prime\prime}+(\frac{v^{2}+2\o}{\a^{2}})\rho-
 \frac{2g}{\a^{2}}\rho^{3}-\frac{2V(\xi)}{\a^{2}}\rho-\frac{2K}{\a^{2}}=0.
 \ee
 Below we consider three different `smart'
Jacobi elliptic potentials \cite{abramowitz} and find exact solutions.\\
\noindent Case(I):- $V(\xi)=-V_{0}{\rm sn}(\xi,m)$. Eq. (3) reads
as \be \rho^{\prime\prime}+(\frac{v^{2}+2\o}{\a^{2}})\rho-
 \frac{2g}{\a^{2}}\rho^{3}+\frac{2V_{0}}{\a^{2}}{\rm
 sn}\rho-\frac{2K}{\a^{2}}=0.
\ee Substituting \be \rho(\xi)=A_{1}+B_{1}{\rm sn}(\xi,m) \ee in
Eq.(4) and equating the coefficients of equal powers of ${\rm
sn}(\xi,m)$ result in relations among the solution parameters
$A_{1}$, and $B_{1}$, and the equation parameters $V_{0}$, $g$,
$K$, $\a$, and $\o$. We find that \bea
\a^{2}=\frac{2\o+v^{2}}{1+m},\\
A_{1}=\frac{V_{0}}{3\a\sqrt{gm}}, ~~~~~~
B_{1}=\sqrt{\frac{m\a^{2}}{g}}. \eea From Eq. (7) it follows that
$V_{0}>0$ and $g>0$ implying the GPE with repulsive nonlinearity.

\noindent Case(II):- $V(\xi)=-V_{0}{\rm cn}(\xi,m)$. Eq. (3) reads
as \be \rho^{\prime\prime}+(\frac{v^{2}+2\o}{\a^{2}})\rho-
 \frac{2g}{\a^{2}}\rho^{3}+\frac{2V_{0}}{\a^{2}}{\rm
 cn}\rho-\frac{2K}{\a^{2}}=0.
\ee Substituting \be \rho(\xi)=A_{2}+B_{2}{\rm cn}(\xi,m) \ee in
Eq.(8) and equating the coefficients of equal powers of ${\rm
scn}(\xi,m)$ result in relations among the solution parameters
$A_{12}$, and $B_{2}$, and the equation parameters $V_{0}$, $g$,
$K$, $\a$, and $\o$. We find that \bea
\a^{2}=\frac{2\o+v^{2}}{1-2m},\\
A_{2}=\frac{V_{0}}{3\a\sqrt{-gm}}, ~~~~~~
B_{1}=\sqrt{\frac{-m\a^{2}}{g}}. \eea From the positivity of
$\a^{2}$ we conclude that ${\rm cn}$ solutions exist for $V_{0}>0$
and $g<0$. The condition $g<0$ corresponds to the GPE with
attractive nonlinearity.

\noindent Case(III):- $V(\xi)=-V_{0}{\rm dn}(\xi,m)$. Eq. (3)
reads as \be \rho^{\prime\prime}+(\frac{v^{2}+2\o}{\a^{2}})\rho-
 \frac{2g}{\a^{2}}\rho^{3}+\frac{2V_{0}}{\a^{2}}{\rm
 dn}\rho-\frac{2K}{\a^{2}}=0.
\ee Substituting \be \rho(\xi)=A_{3}+B_{3}{\rm dn}(\xi,m) \ee in
Eq.(12) and equating the coefficients of equal powers of ${\rm
dn}(\xi,m)$ result in relations among the solution parameters
$A_{3}$, and $B_{3}$, and the equation parameters $V_{0}$, $g$,
$K$, $\a$, and $\o$. We find that \bea
\a^{2}=\frac{2\o+v^{2}}{m-2},\\
A_{3}=\frac{V_{0}}{3\a\sqrt{-g}}, ~~~~~~
B_{3}=\sqrt{\frac{-\a^{2}}{g}}. \eea Here we conclude that ${\rm
dn}$ solutions exist only for $V_{0}>0$ and $g<0$.
\section{Rational solutions}
In order to obtain Lorentzian-type of solutions of Eq. (3) we use
a fractional transform \be \rho(\xi)=\frac{A+Bf^{2}}{1+Df} \ee
where $f$ is the respective Jacobi elliptic functions. Again we
obtain the Lorentzian-type of solutions of Eq. (3) for three
different `smart' potentials.\\
\noindent Case(I):-$V(\xi)=-V_{0}{\rm sn}(\xi,m)$. Eq. (3) reads
as \be \rho^{\prime\prime}+(\frac{v^{2}+2\o}{\a^{2}})\rho-
 \frac{2g}{\a^{2}}\rho^{3}+\frac{2V_{0}}{\a^{2}}{\rm
 sn}\rho-\frac{2K}{\a^{2}}=0.
\ee Substituting \be \rho(\xi)=\frac{A_{4}+B_{4}{\rm
sn}^{2}(\xi,m)}{1+D_{1}{\rm sn}(\xi,m)} \ee in Eq.(17) and
equating the coefficients of equal powers of ${\rm sn}(\xi,m)$
will yield the following consistency conditions.
 \bea
2B_{4}+2A_{4}D^{2}_{1}+\Gamma
A_{4}-\frac{2g}{\a^{2}}A^{3}_{4}-\frac{2K}{\a^{2}}=0,\\
2mB_{4}D^{2}_{1}-\frac{2g}{\a^{2}}B^{3}_{4}=0,\\
6mB_{4}D_{1}+\frac{2V_{0}}{\a^{2}}B_{4}D^{2}_{1}=0,\\
6mB_{4}+B_{4}D^{2}_{1}(\Gamma-m-1)-\frac{6g}{\a^{2}}A_{4}B^{2}_{4}+
\frac{4V_{0}}{\a^{2}}B_{4}D_{1}=0,\\
-2mA_{4}D_{1}+B_{4}D_{1}(2\Gamma-3m-3)+\frac{2V_{0}}{\a^{2}}A_{4}D^{2}_{1}
+\frac{2V_{0}}{\a^{2}}B-\frac{2K}{\a^{2}}D^{3}_{1}=0,\\
-4B_{4}(1+m)+A_{4}D^{2}_{1}(\Gamma-m-1)+\Gamma
B_{4}-\frac{6g}{\a^{2}}A^{2}_{4}B_{4}+
\frac{4V_{0}}{\a^{2}}A_{4}D_{1}-\frac{6K}{\a^{2}}D^{2}_{1}=0,\\
A_{4}D_{1}(2\Gamma+m+1)+\frac{2V_{0}}{\a^{2}}A_{4}-\frac{6K}{\a^{2}}D_{1}=0.
 \eea
 From the above consistency conditions we obtain the following
 relations.
 \bea
A_{4}=\frac{18mK}{3m(m+1)\a^{2}+6m\Gamma\a^{2}-\frac{2V^{2}_{0}}{\a^{2}}},\\
B_{4}=-\frac{3\a^{3}m^{3/2}}{V_{0}g^{1/2}},~~~~~~~~
D_{1}=\frac{-3m\a^{2}}{V_{0}},
 \eea
where $\Gamma=\frac{v^{2}+2\o}{\a^{2}}$. Here, we would like to
emphasize that these results generalize those contained in Ref
\cite{paul}, for nonzero trapping potential. This stems from the
fact that the constant $B$ in expression (18) is nonzero, which
follows from the choice of our `smart' potential in Eq. (1).

\noindent Case(II):-$V(\xi)=-V_{0}{\rm cn}(\xi,m)$. Eq. (3) reads
as \be \rho^{\prime\prime}+(\frac{v^{2}+2\o}{\a^{2}})\rho-
 \frac{2g}{\a^{2}}\rho^{3}+\frac{2V_{0}}{\a^{2}}{\rm
 cn}\rho-\frac{2K}{\a^{2}}=0.
\ee Substituting \be \rho(\xi)=\frac{A_{5}+B_{5}{\rm
cn}^{2}(\xi,m)}{1+D_{2}{\rm cn}(\xi,m)} \ee in Eq.(28) and
equating the coefficients of equal powers of ${\rm cn}(\xi,m)$
will yield the following consistency conditions.
 \bea
 2B_{5}(1-m)+2A_{5}D^{2}_{2}(1-m)+\Gamma
 A_{5}-\frac{2g}{\a^{2}}A^{3}_{5}-\frac{2K}{\a^{2}}=0,\\
-2mB_{45}D^{2}_{2}-\frac{2g}{\a^{2}}B^{3}_{5}=0,\\
-6mB_{5}D_{2}+\frac{2V_{0}}{\a^{2}}B_{5}D^{2}_{2}=0,\\
-6mB_{5}+B_{5}D^{2}_{2}(\Gamma+2m-1)-\frac{6g}{\a^{2}}A_{5}B^{2}_{5}+\frac{4V_{0}}{\a^{2}}B_{5}D_{2}=0,\\
2mA_{5}D_{2}+B_{5}D_{2}(2\Gamma+6m-3)+\frac{2V_{0}}{\a^{2}}A_{5}D^{2}_{2}
+\frac{2V_{0}}{\a^{2}}B_{5}-\frac{2K}{\a^{2}}D^{3}_{2}=0,\\
-4B_{5}(1-2m)+A_{5}D^{2}_{2}(\Gamma+2m-1)+\Gamma
B_{5}-\frac{6g}{\a^{2}}A^{2}_{5}B_{5}+\frac{4V_{0}}{\a^{2}}A_{5}D_{2}-\frac{6K}{\a^{2}}D^{2}_{2}=0,\\
A_{5}D_{2}(2\Gamma-2m+1)+\frac{2V_{0}}{\a^{2}}A_{5}-\frac{6K}{\a^{2}}D_{2}=0.
\eea
From the above consistency conditions we obtain the following
relations \bea
A_{5}=\frac{18mK}{3m(1-2m)\a^{2}+6m\Gamma\a^{2}+\frac{2V^{2}_{0}}{\a^{2}}},\\
B_{5}=\frac{3\a^{3}m}{V_{0}}\sqrt{-\frac{m}{g}},~~~~~~~~
D_{2}=\frac{3m\a^{2}}{V_{0}}.
 \eea

\noindent Case(III):- $V(\xi)=-V_{0}{\rm dn}(\xi,m)$. Eq. (3)
reads as \be \rho^{\prime\prime}+(\frac{v^{2}+2\o}{\a^{2}})\rho-
 \frac{2g}{\a^{2}}\rho^{3}+\frac{2V_{0}}{\a^{2}}{\rm
 dn}\rho-\frac{2K}{\a^{2}}=0.
\ee Substituting \be \rho(\xi)=\frac{A_{6}+B_{6}{\rm
dn}^{2}(\xi,m)}{1+D_{3}{\rm dn}(\xi,m)} \ee in Eq.(39) and
equating the coefficients of equal powers of ${\rm dn}(\xi,m)$
will yield the following consistency conditions.
 \bea
 2A_{6}D^{2}_{3}(m-1)+\Gamma
 A_{6}-\frac{2g}{\a^{2}}A^{3}_{6}-\frac{2K}{\a^{2}}=0,\\
-2B_{6}D^{2}_{3}-\frac{2g}{\a^{2}}B^{3}_{6}=0,\\
-4B_{6}D_{3}+\frac{2V_{0}}{\a^{2}}B_{6}D^{2}_{3}=0,\\
-4B_{6}-2B_{6}D_{3}+B_{6}D^{2}_{3}(\Gamma-m+2)-\frac{6g}{\a^{2}}A_{6}B^{2}_{6}+\frac{4V_{0}}{\a^{2}}B_{6}D_{3}=0,\\
2A_{6}D_{3}+B_{6}D_{3}(2\Gamma-3m+4)+\frac{2V_{0}}{\a^{2}}A_{6}D^{2}_{3}
+\frac{2V_{0}}{\a^{2}}B_{6}-\frac{2K}{\a^{2}}D^{3}_{3}=0,\\
-2B_{6}(m-2)+2B_{6}D_{3}+A_{6}D^{2}_{3}(\Gamma-m+2)+\Gamma
B_{6}-\frac{6g}{\a^{2}}A^{2}_{6}B_{6}+\frac{4V_{0}}{\a^{2}}A_{6}D_{3}-\frac{6K}{\a^{2}}D^{2}_{3}=0,\\
A_{6}D_{3}(2\Gamma+m-2)+\frac{2V_{0}}{\a^{2}}A_{6}-\frac{6K}{\a^{2}}D_{3}=0.
\eea
From the above consistency conditions we obtain the following
relations \bea
A_{6}=\frac{6K}{\a^{2}(2\Gamma+m-2)+\frac{2V^{2}_{0}}{\a^{2}}},\\
B_{6}=\frac{2\a^{2}}{V_{0}}\sqrt{-\frac{\a^{2}}{g}},~~~~~~~~
D_{2}=\frac{2\a^{2}}{V_{0}}.
 \eea
 \begin{figure}[t]
\begin{center}
\includegraphics[width=.45\textwidth]{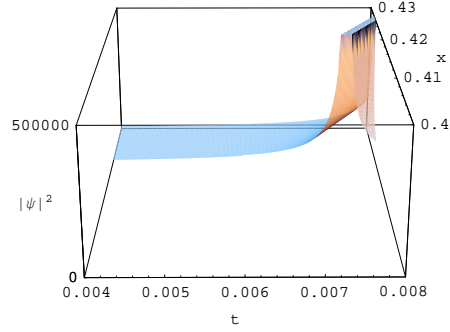}
\caption{Singular solitary wave solution for $\a=1$ and $V_{0}=1$,
and $g=1$.}
\end{center}
\end{figure}
 \subsection{Trigonometric solutions}
 From the consistency conditions that arise from the first two
 `smart' periodic potentials, we conclude that the limit $m=0$ is
 forbidden as the amplitude parameters $A$, $B$, and $D$ will be
 zero. On the other hand, for $V(\xi)=-V_{0}{\rm dn}(\xi,m)$ case,
 only flat background solutions will be possible for $m=0$ limit.
 \subsection{Solitonlike solutions}
 Here, in this subsection, we describe the solitonlike solutions that are obtained from the solutions in
 ${\rm sn}(\xi,m)$ and ${\rm cn}(\xi,m)$ in the limit $m=1$, in detail. In the
 limit $m=1$, $V(\xi)$ becomes an array of well separated
 kink-type of potential barriers: $V(\xi)=-V_{0}{\rm tanh}(\xi)$.
 Then we have the following relations
  \bea
A_{4}=\frac{18K}{6(\Gamma+1)\a^{2}-\frac{2V^{2}_{0}}{\a^{2}}},\\
B_{4}=\frac{3\a^{3}}{V_{0}g^{1/2}},~~~~~~~~
D_{1}=\frac{-3\a^{2}}{V_{0}}.
 \eea
And the strength of the source is
$$K=\frac{V_{0}[6(\Gamma+1)\a^{4}-2V^{2}_{0}]}{108\a^{3}g^{1/2}}\left(\frac{63(\Gamma-2)\a^{4}-
2V^{2}_{0}}{V^{2}_{0}}\right).$$ As a special case, if we set
$\a=1$ and $V_{0}=1$, then we get $A_{4}=\frac{9K}{3\Gamma+2}$,
$B_{4}=3/\sqrt{g}$, and $D_{1}=-3$. This results in a solitonlike
solution \be\rho(\xi)=\frac{A_{4}+B_{4}{\rm
tanh}^{2}(\xi)}{1-3{\rm tanh}(\xi)}. \ee This set corresponds to
the singular solution for repulsive case i.e., $g>0$. The
singularity of the pulse profile may correspond to the beam power
exceeding material breakdown due to self-focussing
\cite{fibich,stolen,boyd,weinstein}. Figure (1) depicts a surface
plot of this solution for the parameter values given in the figure
caption.

Another interesting solitonlike solution is obtained from the
solution in ${\rm cn}(\xi,m)$ for $m=1$. In the limit $m=1$,
$V(\xi)$ becomes an array of well separated
 secant hyperbolic potential barriers: $V(\xi)=-V_{0}{\rm
 sech}(\xi)$.Then we have the following relations
 \bea
A_{5}=\frac{18K}{3(2\Gamma-1)\a^{2}+\frac{2V^{2}_{0}}{\a^{2}}},\\
B_{5}=\frac{3\a^{3}}{V_{0}}\sqrt{-1/g},~~~~~~~~
D_{2}=\frac{3\a^{2}}{V_{0}}.
 \eea
 As a special case, if we set
 $\a=1$, $V_{0}=-6$, $g=-1$ and $K=1/2$ then we get $A_{5}=\frac{9}{6\Gamma+69}$,
$B_{5}=-(1/2)$, and $D_{1}=-(1/2)$. This results in a solitonlike
solution \be\rho(\xi)=\frac{A_{5}+B_{5}{\rm
sech}^{2}(\xi)}{1-0.5{\rm sech}(\xi)}. \ee This set corresponds to
the non-singular solution for attractive case i.e., $g<0$. The
same has been depicted in Fig. 2 for the parameter values given in
in the figure caption.
 \begin{figure}[t]
\begin{center}
\includegraphics[width=.45\textwidth]{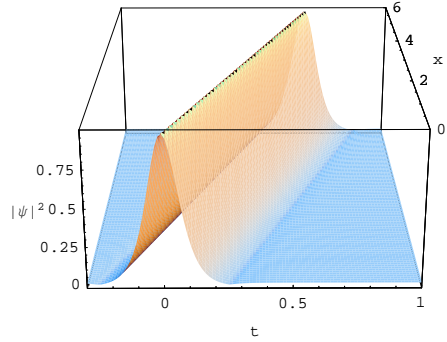}
\caption{Non-singular solitary wave solution for $\a=1$,
$V_{0}=-6$, $g=-1$ and $K=1/2$.}
\end{center}
\end{figure}
 \section{Conclusions}
 In conclusion, we have shown the existence of wide class of exact
 solutions for the modified GP equation in `smart' periodic
 potentials with an external source. The Lorentzian-type of
 solitons are obtained with the aid of a fractional transform. Our
 analysis applies to both attractive and repulsive cases of GP
 equation. Furthermore, our rational solutions generalize those
 contained in Ref~\cite{paul}, for nonzero trapping potential,
 because of our choice of `smart' potential. We hope that these
 potentials may be experimentally realizable, to achieve BEC.
 Although not presented here, the stability of these wide class of
 solutions can be checked using a semi-implicit Crank-Nicholson
 finite difference method \cite{panigrahi}, as the much used numerical techniques
 based on fast Fourier transform (FFT) requires the FFT of the
 source, which is costly.
 \section*{Acknowledgements}
TSR would like to dedicate this paper to the fond memory of his
father Shri.~Thokala Ratna Raju, for his love and encouragement.

\end{document}